\documentclass[superscriptaddress,aps,prx,twocolumn,floatfix]{revtex4-2}
\usepackage{amsmath,amssymb,graphicx,dcolumn,bm,hyperref}
\usepackage{multirow}
\usepackage{xcolor}
\usepackage{orcidlink}

\begin{document}

\title{Microscopic foundation for \textit{orbitronics}: orbital exchange interactions, orbital-dependent transport, and orbital switching}

\author{Hantao Zhang\,\orcidlink{0000-0001-6281-6708}}
\affiliation{Department of Electrical and Computer Engineering, University of California, Riverside, CA 92521, USA}
\affiliation{School of Engineering and Applied Science, The George Washington University, Washington, DC 20052, USA}
\author{Ran Cheng\,\orcidlink{0000-0003-0166-2172}}
\affiliation{Department of Electrical and Computer Engineering, University of California, Riverside, CA 92521, USA}
\affiliation{Department of Physics and Astronomy, University of California, Riverside, CA 92521, USA}
\affiliation{Department of Materials Science and Engineering, University of California, Riverside, CA 92521, USA}

\begin{abstract}
Recent experiments highlighted the essential role of the orbital degrees of freedom underlying current-induced magnetic dynamics. However, a microscopic picture of how orbital angular momenta are transferred between itinerant and localized electrons, and the way it feeds back into transport, remains elusive. Here we develop a theoretical framework for the orbital exchange interactions by generalizing the Anderson-Kondo model to incorporate the orbitals of conduction bands. Illustrating with the $p$-orbitals, we determine the way electrons acquire orbital polarization and its influence on the transport properties, uncovering the orbital counterparts of phenomena analogous to spin filtering and tunnel magnetoresistance. Reciprocally, we demonstrate that orbital-polarized currents can exert torques on localized magnetic moments without having to convert into spin currents, which enables electrical switching of magnetization even when spin polarization becomes inactive. These striking discoveries lay a microscopic foundation for orbitronics and reveal critical mechanisms for orbital manipulations of magnetic order.
\end{abstract}

\maketitle

\section{Introduction}

Manipulation and transport of angular momentum in solids form the conceptual backbone of modern information physics. As a promising alternative to the established paradigm of electronics, spintronics exploits the intrinsic spin of electrons as information carrier. Nevertheless, spintronic operations inevitably rely on dissipative charge currents, because spin polarization is typically generated from charge transport through the relativistic spin-orbit coupling (SOC)~\cite{maekawa2017spin}. This inherent limitation spurred a growing interest in the orbital degree of freedom as a potential carrier of angular momentum, giving rise to the emerging playground of orbitronics~\cite{go2021orbitronics,jo2024spintronics,go2024non,lee2024electric,wang2025orbitronics,cysne2025orbitronics,ando2025orbitronics,fukami2025challenges}.

The orbital angular momentum (OAM) originates from the spatial structure and symmetry of electronic wave functions~\cite{go2018intrinsic,jo2018gigantic,go2024non}, which prevails even in a pure non-relativistic picture. Consequently, the OAM remains well-defined in materials with weak or negligible SOC, suggesting the possibility of long-range transmission of angular momenta that circumvents the physical limit of spin diffusion~\cite{go2023long,hayashi2023observation,gao2024control}. Motivated by this simple principle, substantial progress has recently been made in developing orbitronics into a field bearing a level of understanding similar to that of spintronics. For example, theoretical studies have shown that momentum-space orbital textures can enable transverse orbital currents driven by an applied electric field, leading to the orbital Hall effect in centrosymmetric and light-element materials~\cite{zhang2005intrinsic,bernevig2005orbitronics,kontani2009giant,tanaka2008intrinsic,go2018intrinsic,jo2018gigantic,baek2021negative}. Complementary mechanisms such as the orbital Edelstein effect can generate nonequilibrium orbital polarization at interfaces~\cite{park2011orbital,go2017toward,johansson2021spin,go2021orbital}, while ultrafast optical excitation has been demonstrated to induce orbital angular momentum on femtosecond timescales~\cite{adamantopoulos2024orbital,adamantopoulos2024spin,adamantopoulos2025light}. These predictions have been followed by a rapidly growing body of experimental work reporting orbital-current generation across a wide range of material platforms~\cite{an2016spin,ding2020harnessing,kim2021nontrivial,lee2021orbital,sala2022giant,choi2023observation,lyalin2023magneto,el2023observation,seifert2023time,hayashi2024observation,xu2024orbitronics}.

Regarding the detection of OAM, both theories and experiments suggest that orbital currents can propagate over long distances in non-magnetic materials and subsequently exert torques on adjacent magnetic layers~\cite{hayashi2023observation,go2023long,ning2024phenomenology,tang2024role,gao2024control,gao2025nonlocal,ding2025generation}, rendering detectable effects with an efficiency comparable to or exceeding that of conventional spin–orbit torques~\cite{chen2018giant,go2020orbital,zheng2020magnetization,ding2020harnessing,lee2021efficient,kim2021nontrivial,lee2021orbital,li2023giant,bose2023detection,ning2025orbital}. However, in essentially all existing realizations of orbital torques, the non-equilibrium OAM does not directly interact with local magnetic order but necessarily converts into spin angular momentum via the SOC; hence the measured signal is ultimately attributed to a spin-orbit torque. In other words, what has been experimentally accessible does not appear to be a genuine orbital torque, but rather spin torques stemming from orbital currents by virtue of the SOC. This situation exposes a serious conceptual dilemma. Although the primary motivation for orbitronics is to move beyond SOC-dominated effects, the SOC appears to be indispensable for the detection and application of orbital currents. Therefore, a fundamental question arises: are there direct interactions between itinerant OAM and localized orbitals without invoking the SOC, which enables orbital-dependent transport and orbital torques obviating spin currents as a mediator?

This question prompts us to hold back from orbitronics for a moment and reflect on the microscopic origin of spin torques, or more broadly, spin transfer between transport electrons and local magnetic moments. Spin conservation, while providing a general principle, is inadequate by itself to provide a closed-form interaction governing the spin transfer processes. What serves the purpose is the exchange interaction that connects the itinerant and localized electrons, which can be traced back to the Anderson-Kondo interaction~\cite{anderson1961localized,schrieffer1966relation,KONDO1970183}. In a spintronic context, this exchange interaction is often simplified into an effective s-d coupling $J_{sd}\bm{\sigma}\cdot\bm{M}$ in the real space~\cite{maekawa2017spin}, where $\bm{\sigma}$ is the spin of the itinerant non-magnetic $s$ electrons and $\bm{M}$ is the magnetic moment, while the momentum-dependence of $J_{sd}$ is ignored. In one way or another, physical quantities pertaining to the transfer of spin angular momenta (such as spin torques, spin magnetoresistance, spin-mixing conductance, etc.) are microscopically originating from this interaction. However, such interaction becomes elusive when it comes to the transfer of OAM. Several works simply assume an interaction of the form $\bm{L}\cdot \bm{M}$, by analogy with spin exchange~\cite{chen2024dzyaloshinskii,sohn2024dyakonov,han2025orbital,lee2026orbital,sohn2024dyakonov}, which has never been rigorously justified. More seriously, this form of coupling obscures a fundamental difficulty unique to OAM. The component $L_x M_x + L_y M_y$ in $\bm{L}\cdot \bm{M}$ can be expressed as $L^+ M^- + L^- M^+$ (where $\pm$ stands for the raising and lowering operators), which implies the conservation of OAM similar to its spin counterpart. However, unlike spin angular momentum, OAM is generally not conserved even in the absence of SOC~\cite{fang2015ab,lee2023effect,lee2024electric}. This inconsistency indicates that the form and origin of orbital exchange interactions must be reconsidered at a more fundamental level. Notably, a rigorous derivation of orbital exchange interactions requires a multi-electron framework~\cite{solidstatphys}, whereas most of the existing theories are based on single-electron pictures. 

In this work, we formulate a microscopic theory of the orbital exchange interaction between itinerant and localized electrons by extending the Anderson–Kondo framework to include orbital degrees of freedom, which does not explicitly rely on the SOC. As a proof-of-concept study, we only
consider the simplest $p$-orbitals to establish the essential physical picture without accounting material details, while the formalism itself can be generalized to more realistic scenarios (with the help of first-principle calculations). However, even for $p$-orbitals, the orbital exchange interaction turns out to be appreciably more complicated than the naive form of $\bm{L}\cdot \bm{M}$. Based upon the derived orbital exchange interaction, we investigate the generation of orbital-polarized current and orbital-dependent magnetoresistance (MR) that are analogous to the spin filtering and the spin MR, respectively. Then, we study the reciprocal effects of the orbital-dependent transport enabled by the same orbital exchange interaction, where we show that an orbital current can directly exert torques on, hence switching, local orbital and spin magnetization. Our theory timely lays a microscopic foundation for exploring the orbital counterparts of various transport phenomena known in spintronics, inviting future inquiries in the thriving field of orbitronics.

\section{Model}
\begin{figure}[t]
\centering
\includegraphics[width=0.8\linewidth]{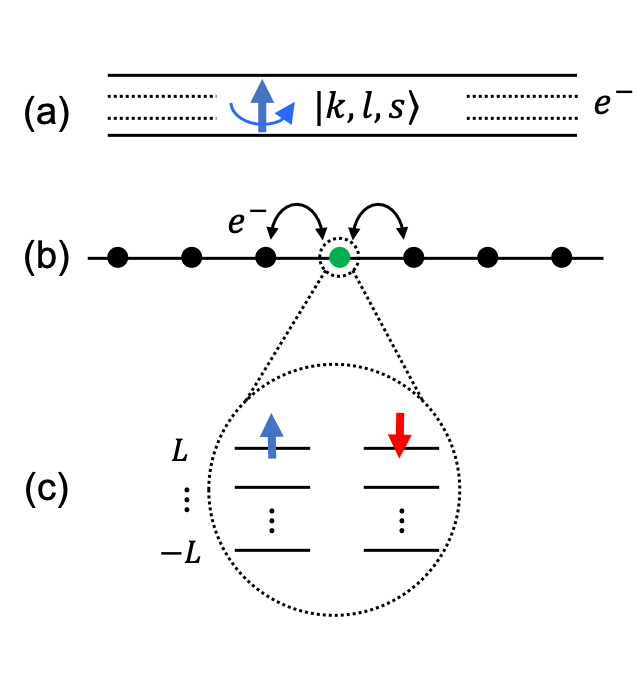}
\caption{Schematic illustration of the model system. The Hamiltonian consists of three parts: (a) conduction band where electrons are labeled by crystal momentum $k$, OAM $\ell$, and spin $s$; (b) electron hopping between neighboring atoms; and (c) localized states with the orbital quantum number ranging from $-L$ to $L$.}
\label{fig:model}
\end{figure}

Generalizing the Anderson–Kondo model, we consider a physical system in which individual magnetic atoms, with multiple orbital and spin states, are embedded in a fermi sea as illustrated in Fig.~\ref{fig:model}. Our analysis focuses on the interaction between itinerant electrons at the Fermi level and the localized electrons occupying the atomic energy states [Fig.~\ref{fig:model}(a)]. Electrons can hop between neighboring atoms, changing the electron occupation of both the localized atomic states and the conduction band, as illustrated in Fig.~\ref{fig:model}(b). Without being restricted by material details, here we demonstrate the essential physical picture in the simplest non-trivial case, namely $p$-orbitals in the localized states, alongside the spin degrees of freedom [Fig.~\ref{fig:model}(c)]. It should be noted that we only consider the intra-atomic contribution to the OAM of itinerant electrons under the atomic center approximation~\cite{cysne2022orbital,pezo2022orbital,pezo2023orbital} while excluding the interatomic contribution. Here, we do not consider $s$-orbitals because without orbital hybridization or interatomic contribution they simply cannot carry any OAM in a metal~\cite{busch2023orbital}. Our formalism is technically generalizable to $d$, $f$, or hybridized orbitals by extending the basis, but the results will be significantly more sophisticated, which goes against our purpose in building a microscopic but intuitive understanding of the orbital exchange interaction. To ensure compatibility with the localized states, the conduction band is also assigned the orbital and spin quantum numbers. Similar to the orbital-free Anderson model, we assume that the coupling between the itinerant and localized states is much smaller than the Coulomb repulsion and the Fermi energy, which will be treated as a perturbation.

With the above considerations, we construct the effective Hamiltonian by including orbital hopping and a multi-orbital Hubbard interaction into the Anderson Hamiltonian:
\begin{subequations}
\label{eq:Hamiltonian}
\begin{align} 
     H&=H_{c} + H_{c-i} + H_{i}, \label{eq:total_H} \\ H_{c}&=\sum_{k,\ell,s}\epsilon_{k,\ell,s}c^{\dagger}_{k,\ell,s} c_{k,\ell,s}, \label{eq:conduct_H} \\
    H_{c-i}&=\sum_{\bm{d},\ell,L,s,\sigma}t^{a}_{\ell,L,\bm{d}} c^{\dagger}_{\ell,s,\bm{d}}c_{L,\sigma}\delta_{s,\sigma} + c.c., \label{eq:hopping_H} \\
    H_{i}= \sum_{L,\sigma}&\epsilon_{L,\sigma}n_{L,\sigma}+\sum_{L,\sigma;L',\sigma'} U_{L,\sigma;L',\sigma'}n_{L,\sigma}n_{L',\sigma'}, \label{eq:impurity_H}
\end{align}
\end{subequations}
where $H_{c}$, $H_{c-i}$, and $H_{i}$ describe, respectively, the conduction band of electrons, the hopping between neighboring atoms, and the atomic energy including Coulomb repulsions for multi-occupations. The quantum numbers $k$, $\ell$, and $s$ denote the momentum, OAM, and spin of the conduction electrons, with $\epsilon_{k,\ell,s}$ their energy. For the localized states, $L$ and $\sigma$ label the orbital and spin, respectively, with $\epsilon_{L,\sigma}$ their single-particle energies. The operators $c^{\dagger}$ ($c$) create (annihilate) electrons, $\bm{d}$ is the bonds between neighboring atoms, $t^{a}_{\ell,L,\bm{d}}$ is the atomic-scale hopping amplitude, $n$ is the number operator, and $U_{L,\sigma;L',\sigma'}$ is the on-site Hubbard interaction. The Kronecker delta symbol $\delta_{s,\sigma}$ enforces conservation of spin angular momentum. As mentioned in the previous paragraph, $\ell$ and $L$ range from $-1$ to $+1$, corresponding to the range of $L_{z}$ of $p$-orbital states. Our formalism is in principle applicable to more realistic orbitals, such as $d$, $f$, and hybridized $s-p$, since $\ell$ and $L$ are just labels of the orbital degrees of freedom, and $t^{a}_{\ell,L,\bm{d}}$ can be determined by material details. In our model, Hund's splitting is very weak such that different $p$-orbitals on an atom are nearly degenerate, which is a valid approximation in the long wavelength regime~\cite{go2023long}, meaning $\epsilon_{L,\sigma} \approx \epsilon_{0}$. Furthermore, since only conduction electrons on the Fermi surface contribute to transport, $\epsilon_{k,\ell,s}\approx\epsilon_F$ for all $\ell$ and $s$, where $\epsilon_{F}$ is the Fermi energy.

Next, we determine the hopping matrix. In the original Anderson–Kondo model, the hopping matrix is diagonal in the spin space because of spin conservation. However, this intuition does not directly extend to the orbital degrees of freedom, as previous works have indicated that inter-orbital hopping can break the conservation of OAM~\cite{fang2015ab,lee2023effect}. Therefore, besides the simplest case of diagonal orbital hopping matrix, we need to check the question in a specific context. In our toy model, we employ atomic orbital wavefunctions rather than Wannier functions as our basis, so no \textit{ab initio} calculations are required. To simplify Eq.~\eqref{eq:hopping_H}, we reduce the problem to the hopping between two isolated atoms:
\begin{align} \label{eq:atomic_hopping}
    t^{a}_{\ell,L,\bm{d}}&=\left<\ell, \bm{R} + \bm{d} \right|\left[-\frac{\hbar^{2}}{2m_{e}}\nabla^{2}\right. \notag \\
    &\qquad \left.-\frac{N_{1}e^{2}}{|\bm{r}-\bm{R}|} -\frac{N_{2}e^{2}}{|\bm{r}-\bm{R}-\bm{d}|}\right] \left|L, \bm{R} \right>,
\end{align}
where $\ell$, $L$ represent the OAM of the electron wavefunctions in the itinerant and localized states, and $N_{i}$ ($i=1,2$) is the net number of electrons in each nucleus. $\bm{R}$ and $\bm{R}+\bm{d}$ indicate the atomic positions, while $\bm{r}$ is the electron position, $m_{e}$ is the electron mass, and $e$ is the electron charge. Because $t^{a}_{\ell,L,\bm{d}}$ depends on the relative orientation between $\bm{d}$ and the orbital quantization axis (taken to be $z$ axis), we consider two representative cases to draw the physical picture. When $\bm{d}\parallel z$, corresponding to an individual atom in a one-dimensional atomic chain, the hopping matrix is indeed diagonal, $t^{a}_{\ell,L,\bm{d}} = t\delta_{\ell,L}$. However, if $\bm{d}$ lies along the $[111]$ direction, corresponding to an individual atom at the center of a body-centered cubic cell, the hopping matrix becomes
\begin{equation} \label{eq:hopping_M_bcc}
t^{a}_{\ell,L,\bm{d}} = 
    \begin{pmatrix}
0 & -t+\mathrm{i} t & \sqrt{2}\mathrm{i}t\\
-t-\mathrm{i}t & 0 & t-\mathrm{i}t \\
-\sqrt{2}\mathrm{i}t & t+\mathrm{i}t & 0
\end{pmatrix},
\end{equation}
where $\mathrm{i}^{2}=-1$, $t$ is a hopping constant and the basis is ${\left|L=1\right>, \left|L=0\right>, \left|L=-1\right>}$ for $p$-orbitals. It is evident that OAM is conserved when $\bm{d}$ is parallel to $z$, but not otherwise. This distinction follows from the local rotational symmetry. Eq.~\eqref{eq:atomic_hopping} is invariant under rotations about the $z$ axis when $\bm{d}\parallel z$; but changes under rotation if $\bm{d}$ deviates from the axis. Since $t^{a}_{\ell,L,\bm{d}}$ depends on the crystal environment, different lattices (e.g., simple cubic or face-centered cubic) would yield different hopping matrices~\cite{fang2015ab,lee2023effect}. Finally, we note that Eq.~\eqref{eq:atomic_hopping} also applies to other orbital hoppings, such as $s$–$d$ and $p$–$d$, but these cases go beyond the scope of the present discussion.

With the obtained hopping matrix, we need to perform a Fourier transformation $c_{\ell,s,\bm{d}}=\sum_{\bm{k}}c_{k, \ell,s}e^{i\bm{k}\cdot\bm{d}}$ in Eq.~\eqref{eq:hopping_H}. Under the long wavelength approximation~\cite{noce2006periodic},
$\sum_{\bm{d}}t^{a}_{\ell,L,\bm{d}}e^{i\bm{k}\cdot\bm{d}}\approx Zt^{a}_{\ell, L, \bm{d}}$ where $Z$ is the coordination number. Then, the hopping Hamiltonian becomes
\begin{align} \label{eq:hopping_H_k}
     H_{c-i}&=\sum_{k,\ell,L,s,\sigma}Zt^{a}_{\ell,L,\bm{d}} c^{\dagger}_{k,\ell,s}c_{L,\sigma}\delta_{s,\sigma} + c.c. \notag \\
     &= \sum_{k,\ell,L,s,\sigma}t_{\ell,L} c^{\dagger}_{k,\ell,s}c_{L,\sigma}\delta_{s,\sigma} + c.c., 
\end{align}
where $t_{\ell,L}$ is the effective hopping matrix. In the long wavelength limit, Eq.~\eqref{eq:impurity_H} is effectively in the Fourier space already, with $k\approx0$.

The last essential ingredient is the multi-orbital Hubbard interaction in Eq.~\ref{eq:impurity_H}. We adopt the on-site interaction matrix $U_{L,\sigma;L',\sigma'}$ as~\cite{daghofer2008model, dagotto2011properties}
\begin{equation} \label{eq:Hubbard_U}
    U_{L,\sigma;L',\sigma'} =
    \begin{cases}
        U, \text{if } L = L', \sigma=-\sigma'\\
        U'+J, \text{if } L \neq L', \sigma=-\sigma' \\
        U'-J, \text{if } L \neq L', \sigma=\sigma' 
    \end{cases},
\end{equation}
where $U$ and $U'$ are orbital-selective Hubbard energies and $J$ is the spin-exchange energy. These parameters satisfy $U'-J<U'+J<U$ and $\epsilon_{0} + U'-J>\epsilon_{F}$. The first constraint reflects Hund’s rules, which favor parallel spin alignment, while the second constraint parallels the condition involving $\epsilon_{0}$, $U$, and $\epsilon_{F}$ in the orbital-free Anderson Hamiltonian. Previous studies also include a pair-hopping term, such as $c^{\dagger}_{L,\sigma}c^{\dagger}_{L,-\sigma}c_{L',-\sigma}c_{L',\sigma}$, with an amplitude proportional to $J$~\cite{daghofer2008model, dagotto2011properties}. However, in our model, changes of the electron occupation are primarily driven by interactions with conduction electrons, and $J$ is typically smaller than both $U$ and $U'$. Consequently, the pair-hopping contribution plays only a secondary role~\cite{lin2021origin}, which will be ignored for simplicity.

\section{Effective Orbital Exchange Interaction} \label{sec:effective_orbital_exchange}
Equipped with the model Hamiltonian, we next need to project out the high-energy many-body states in the Hilbert space. This procedure, originally exploited to derive the effective Kondo Hamiltonian (the $s$-$d$ exchange) from the Anderson model, is followed here with a multi-orbital basis in mind. The result is a low-energy effective Hamiltonian acting within the degenerate ground-state subspace, which is identified as the effective orbital exchange interaction.

\subsection{Schrieffer–Wolff transformation}
An essential theoretical tool we need is the Schrieffer–Wolff (SW) transformation~\cite{bravyi2011schrieffer,solidstatphys}. Consider a Hamiltonian $H = H_{0} + \Delta H$ where $H_{0}$ is the unperturbed part and $\Delta H$ is a first-order perturbation that couples subspaces of different energies in the Hilbert space. Our aim is to construct an effective Hamiltonian $H_{\text{eff}}$ that acts only in the degenerate ground-state subspace, without mixing with the excited states. Since any first-order perturbation necessarily couples different subspaces, $H_{\text{eff}}$ must arise from the second order perturbation in $\Delta H$.

Consider the effective Hamiltonian as related to the total Hamiltonian by a unitary transformation, $H_{\text{eff}}=e^{S}He^{-S}$, where $S$ is on the same order as $\Delta H$. Using the Lie group homomorphism~\cite{liegroupbook} and truncating to the second order, we obtain
\begin{align}
 H_{\text{eff}} = H + [S,H] +\frac12[S,[S,H_{0}]]+\cdots
\end{align}
Requiring $[S,H_{0}] = -\Delta H$ then cancels the first-order perturbation, leaving $H_{\text{eff}} = H_{0}+[S,\Delta H]/2$. Let ${|\mu\rangle}$ be the eigenstates of $H_{0}$ with $H_{0}|\mu\rangle = E_{\mu}|\mu\rangle$, so $S=\sum_{\mu,\nu} \left|\mu\right>\left<\nu\right| \left< \mu \right| \Delta H \left| \nu \right>/(E_{\mu}-E_{\nu})$, where the sum excludes terms with $E_{\mu} = E_{\nu}$. Substituting $S$ into the expression for $H_{\text{eff}}$ gives
\begin{align}
\label{eq:SW_Heff}
    \Delta H_{\rm eff} &= \frac{1}{2}\sum_{\xi,\eta}\sum_{\lambda \neq \xi,\eta} \left|\xi\right>\left<\eta\right| \left< \xi \right| \Delta H \left| \lambda \right> \left< \lambda \right| \Delta H \left| \eta \right> \notag \\
    &\qquad \times \left( \frac{1}{E_{\xi} - E_{\lambda}} + \frac{1}{E_{\eta}-E_{\lambda}} \right).
\end{align}
Since $|\xi\rangle$ and $|\eta\rangle$ both lie in the ground-state subspace, $\Delta H_{\text{eff}}$ acts entirely within this subspace, yielding the desired low-energy effective Hamiltonian.

\subsection{Effective orbital Hamiltonian}
We apply the SW transformation to Eq.~\eqref{eq:Hamiltonian} by considering $H_{c-i}$ in Eq.~\eqref{eq:hopping_H_k} as the perturbation $\Delta H$. The validity of this procedure is ensured by the well-separated energy levels associated with different occupations of the localized states on magnetic atoms. Furthermore, the condition $|t^{a}_{\ell,L,\bm{d}}| \ll U, U'$ guarantees that the subspaces spanned by degenerate many-body states remain weakly coupled. Without losing the essential physics, we restrict our discussion to the simple case $t^{a}_{\ell,L,\bm{d}} = t\delta_{\ell,L}$ (e.g., $\bm{d}\parallel z$ in a 1D chain, so $t^{a}_{\ell,L,\bm{d}} = t\delta_{\ell,L}$ as mentioned previously). In real materials, the hopping matrix can be quite complicated, which calls for \textit{ab initio} calculations. Such restriction, however, will be lifted when we discuss orbital transport and orbital torques in the following sections, where we will instead adopt Eq.~\eqref{eq:hopping_M_bcc}.

Unlike the original spin-only Anderson model, which involves only two degenerate states with a single electron occupying one of them, our model permits double and triple occupations. As a result, a key feature of our formalism is that the effective orbital Hamiltonian explicitly depends on the electron occupation of the localized states. For the single occupation case, after some tedious calculations, we obtain the effective Hamiltonian as

\begin{subequations} \label{eq:effect_H_sing}
\begin{align}
\Delta & H_{\rm eff} = -|t|^{2} \sum_{k} \Delta H_{\text{eff},s}^{k} + \Delta H_{\text{eff},o}^{k} + \Delta H_{\text{eff},so}^{k} \nonumber \\
&+ \Delta H_{\text{eff},c}^{k}, \label{eq:H_eff} \\
\Delta & H_{\text{eff},s}^{k} = 4\left( \frac{1}{\epsilon_{0}-\epsilon_{F}} + \frac{1}{\epsilon_{F}-\epsilon_{0}-U}\right) & \nonumber \\
& \times  \left( s^{+}_{k}\sigma^{-} + s^{-}_{k}\sigma^{+} \right), \label{eq:H_eff_s} \\
 \Delta & H_{\text{eff},o}^{k} = \frac{1}{4}\left( \frac{1}{\epsilon_{0}-\epsilon_{F}} + \frac{1}{\epsilon_{F}-\epsilon_{0}-U'+J}\right) \nonumber \\
 &\times \left[  
    2\left( \ell^{+}_{k}L^{-}+ \ell^{-}_{k}L^{+} \right) + \left( \ell^{+}_{k}L^{-}\right)^{2} + \left(\ell^{-}_{k}L^{+} \right)^{2}\right], \label{eq:H_eff_o} \\
\Delta & H_{\text{eff},so}^{k} = \left( \frac{1}{\epsilon_{0}-\epsilon_{F}} + \frac{1}{\epsilon_{F}-\epsilon_{0}-U'-J}\right) \nonumber \\
&\times \left[  
    2\left( \ell^{+}_{k}L^{-} + \ell^{-}_{k}L^{+} \right)  + \left( \ell^{+}_{k}L^{-}\right)^{2} + \left(\ell^{-}_{k}L^{+} \right)^{2}\right] \nonumber \\
&\times \left( s^{+}_{k}\sigma^{-} + s^{-}_{k}\sigma^{+} \right),\label{eq:H_eff_so} \\      
\Delta & H_{\text{eff},c}^{k} = \frac{1}{\epsilon_0 - \epsilon_F} + \frac{1}{\epsilon_{F}-\epsilon_{0}-U} + \frac{2}{\epsilon_F - \epsilon_0 -U' - J} \nonumber \\
&+ \frac{2}{\epsilon_F - \epsilon_0 -U' + J}, \label{eq:H_eff_c}
\end{align}
\end{subequations}

where $s^{\pm}_{k}$ ($\sigma^{\pm}$) and $\ell^{\pm}_{k}$ ($L^{\pm}$), absorbing a factor of $1/\hbar$, are the spin and OAM ladder operators for the itinerant electrons (localized electrons). The overall minus sign in Eq.~\eqref{eq:H_eff}, compared to Eq.~\eqref{eq:SW_Heff}, originates from the exchange of conduction and local electrons~\cite{solidstatphys}.

\begin{figure}[t]
\centering
\includegraphics[width=\linewidth]{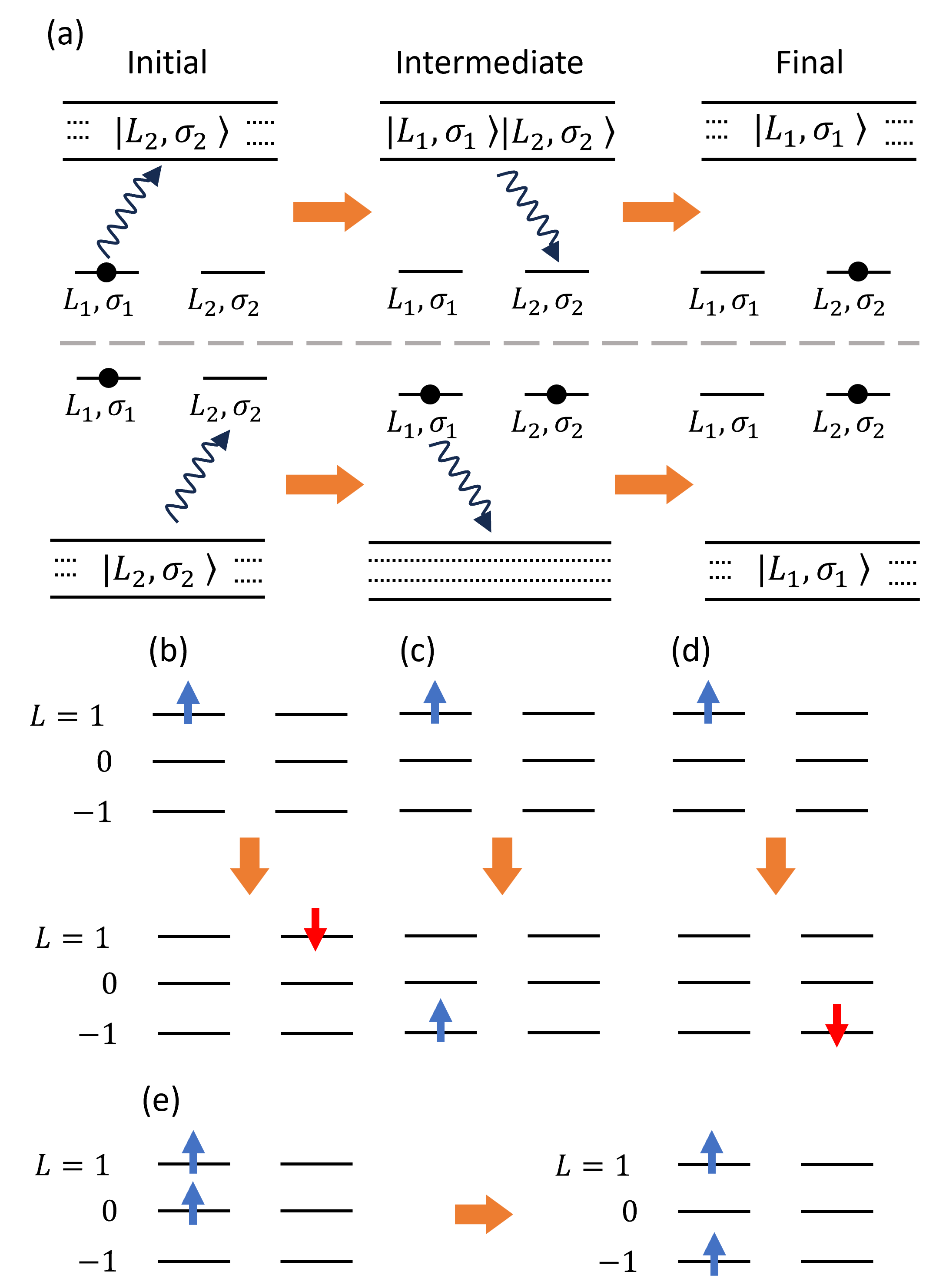}
\caption{ (a) The initial, intermediate and final states during the electron exchange process. We only plot the two relevant states and ignore other states for simplicity. Upper and lower panels show different intermediate states during the exchange process. The initial and final local states involving in the (b) spin-only exchange, (c) orbital-only exchange and (d) spin-orbital exchange, for single occupation, and (e) orbital-only exchange for double occupation. We omit the change of conduction electrons for simplicity as it can be inferred from the change of local states.}
\label{fig:exchange_process}
\end{figure}

Figure~\ref{fig:exchange_process}(a) illustrates the physical picture associated with the effective Hamiltonian Eq.~\eqref{eq:effect_H_sing}, where the spin and orbital exchange processes share a similar picture. Suppose that initially an electron occupies the state $L,\sigma$, making the system a single occupation. Under the perturbation $H_{c-i}$, two distinct processes can occur. In the first case, the local electron hops into the conduction band, leaving the local site temporarily empty [Fig.~\ref{fig:exchange_process}(a), upper panel]. In the second case, a conduction electron hops into the local site, creating a double occupation [Fig.~\ref{fig:exchange_process}(a), lower panel]. Both processes are completed when a conduction (local) electron hops back to the local site (conduction band), returning the system to a singly occupied configuration. The final state then contains an electron in $L',\sigma'$. These two types of virtual processes, distinguished by their intermediate occupations, yield different energy denominators and thereby contribute to the amplitudes of the effective exchange interactions. Specifically, the process in the upper panel corresponds to the terms proportional to $1/(\epsilon_{0}-\epsilon_{F})$ in Eqs.~\eqref{eq:H_eff_s}, \eqref{eq:H_eff_o}, and \eqref{eq:H_eff_so}, while the process in the lower panel gives rise to other prefactors.

The effective Hamiltonian in Eq.~\eqref{eq:effect_H_sing} naturally separates into four components, Eqs.~\eqref{eq:H_eff_s}–\eqref{eq:H_eff_c}.
\begin{itemize}
    \item Spin-only exchange ($\Delta H^{k}_{\text{eff},s}$): This term describes interactions purely between spin degrees of freedom, independent of orbital states. It reproduces the standard Kondo Hamiltonian and has been widely studied. The corresponding occupations of the localized states are shown in Fig.~\ref{fig:exchange_process}(b).
    \item Orbital-only exchange ($\Delta H^{k}_{\text{eff},o}$): This term captures exchange interactions mediated solely by the orbital degrees of freedom. While $\ell^{+}_{k}L^{-}+\ell^{-}_{k}L^{+}$ resembles the spin-only exchange form, the $(\ell^{+}_{k}L^{-})^{2} + (\ell^{-}_{k}L^{+})^{2}$ term is a more complicated form of interaction involving OAM. For $p$ orbitals, the maximum power of ladder operators is 2 (connecting $L=1$ and $L=-1$), while more complicated terms may appear for $d$ or $f$ orbitals. The local occupations before and after this orbital-only exchange process are shown in Fig.~\ref{fig:exchange_process}(c).
    \item Spin–orbital exchange ($\Delta H^{k}_{\text{eff},so}$): This term describes intertwined spin–orbital exchange, despite the absence of explicit SOC in our model. It simultaneously changes both the spin and orbital quantum numbers, as illustrated in Fig.~\ref{fig:exchange_process}(d). To the best of our knowledge, such a contribution has not been explicitly considered in previous studies.
    \item Constant shift ($\Delta H^{k}_{\text{eff},c}$): This term modifies the overall energy without changing the spin or orbital states.
\end{itemize}

In the spin-only case (Anderson–Kondo model), combining $\Delta H^{k}_{\text{eff},s}$ and $\Delta H^{k}_{\text{eff},c}$ yields the familiar form $\bm{s}_{k}\cdot\bm{\sigma}$. While a similar simplification holds here for the spin channel, the more complex structures of $\Delta H^{k}_{\text{eff},o}$ and $\Delta H^{k}_{\text{eff},so}$ prevent a straightforward reduction of Eq.~\eqref{eq:effect_H_sing} to $\bm{\ell}_{k}\cdot\bm{L}$. This means that the neat form of the $s$–$d$ exchange cannot be naively generalized to orbital interactions.

Finally, the relative strengths of these interactions differ because their amplitudes depend on $U$, $U'+J$, and $U'-J$, as dictated by Hund’s rules. Under the condition $U'-J < U'+J < U$, the orbital-only exchange ($\Delta H^{k}_{\text{eff},o}$) dominates, while the spin-only exchange ($\Delta H^{k}_{\text{eff},s}$) is the weakest. If one disregards Hund’s rules, the three exchange amplitudes in Eqs.~\eqref{eq:effect_H_sing}(a)-(c) would become the same, but the three terms are still physically distinguishable and remain in the theory.

We now consider the case of double occupation, for which the effective Hamiltonian is derived as

    \begin{subequations} \label{eq:effect_H_doub}
        \begin{align}
            &\Delta H_{\rm eff} = -|t|^{2} \sum_{k} \Delta H_{\text{eff},o}^{k} + \Delta H_{\text{eff},c}^{k}, \label{eq:H_eff_doub}\\
            &\Delta H_{\text{eff},o}^{k} \nonumber \\
            &= \frac{1}{4}\left(\frac{1}{U'-J+\epsilon_{0}-\epsilon_{F}} + \frac{1}{\epsilon_{F} - \epsilon_{0} - 2U' + 2J} \right) \nonumber \\
             &\times \left[  2\left( \ell^{+}_{k}L^{-}+ \ell^{-}_{k}L^{+} \right) + \left( \ell^{+}_{k}L^{-}\right)^{2} + \left(\ell^{-}_{k}L^{+} \right)^{2}\right] , \label{eq:H_eff_o_doub} \\
    &\Delta H_{\text{eff},c}^{k} = \frac{2}{U'-J+\epsilon_{0}-\epsilon_{F}} + \frac{1}{\epsilon_{F}-\epsilon_{0} - 2U'-2J}  \nonumber \\  
    &+ \frac{1}{\epsilon_{F}-\epsilon_{0} - 2U'+2J} + \frac{2}{\epsilon_{F}-\epsilon_{0} - U'-U-J}, \label{eq:H_eff_c_doub}
        \end{align}
    \end{subequations}
where only the orbital-only exchange interaction survives, as illustrated in Fig.~\ref{fig:exchange_process}(e); whereas spin-only and spin–orbital exchanges are completely absent. The reason is rooted in the multi-orbital Hubbard model of Eqs.~\eqref{eq:impurity_H} and \eqref{eq:Hubbard_U}. The ground states consist of two electrons with parallel spins occupying different orbitals. If a single localized electron is exchanged with a conduction electron while flipping its spin, the resulting state lies outside the ground-state subspace and is therefore excluded from the effective Hamiltonian. If two localized electrons are exchanged with two conduction electrons so that both spins flip, the final state will remain in the ground-state manifold, but the corresponding amplitude is proportional to $|t|^{4}$, which is a higher-order process than Eq.~\eqref{eq:H_eff_doub}. Since the lowest-order effective Hamiltonian from Eq.~\eqref{eq:SW_Heff} only describes dynamics within the ground-state subspace, neither spin-flipping process is captured by Eq.~\eqref{eq:H_eff_doub}. Comparing Eqs.~\eqref{eq:H_eff_o_doub} and \eqref{eq:H_eff_o}, we see that both the forms and amplitudes of orbital-exchange terms change depending on the local occupation. Therefore, the form of the effective exchange interaction, whether for spin, orbital, or spin–orbital, sensitively depends on the local occupation.

In the case of triple occupation, the ground-state configuration of the local site is either all spin up or all spin down, which are connected via higher order transition processes. Regarding the lowest order process considered here, there will not be any orbital or spin exchange interactions; only $\Delta H^{k}_{\text{eff},c}$ exists. For occupations with four and five electrons, they are related to the double and single occupations through the particle-hole transformation. The resulting $\Delta H_{\text{eff}}$ share similar forms as those of double and single occupations, although the prefactors are different due to different energies of the intermediate states.

\section{Orbital Transport}
In this section, we examine how the spin and orbital degrees of freedom of the localized states affect the electron transport. Their impact manifests in two ways: the spin–orbital filter effect (SOFE) and the spin–orbital magnetoresistivity (SOMR). SOFE describes how individual atoms can polarize an initially unpolarized current, while SOMR characterizes the resistivity of an orbital-polarized current as it passes by the magnetic atoms. They bring about highly non-trivial generalizations to their spin counterparts. To simplify the notation, the resistivity will be conveniently expressed in terms of conductivity, and one should bear in mind that $\rho_{xx}=\sigma_{xx}/(\sigma_{xx}^2+\sigma_{xy}^2) = 1/\sigma_{xx}$ as our system does not have Hall conductivity $\sigma_{xy}$.

\subsection{Transport and Tunneling Mechanisms}

\begin{figure}
\centering
\includegraphics[width=1\linewidth]{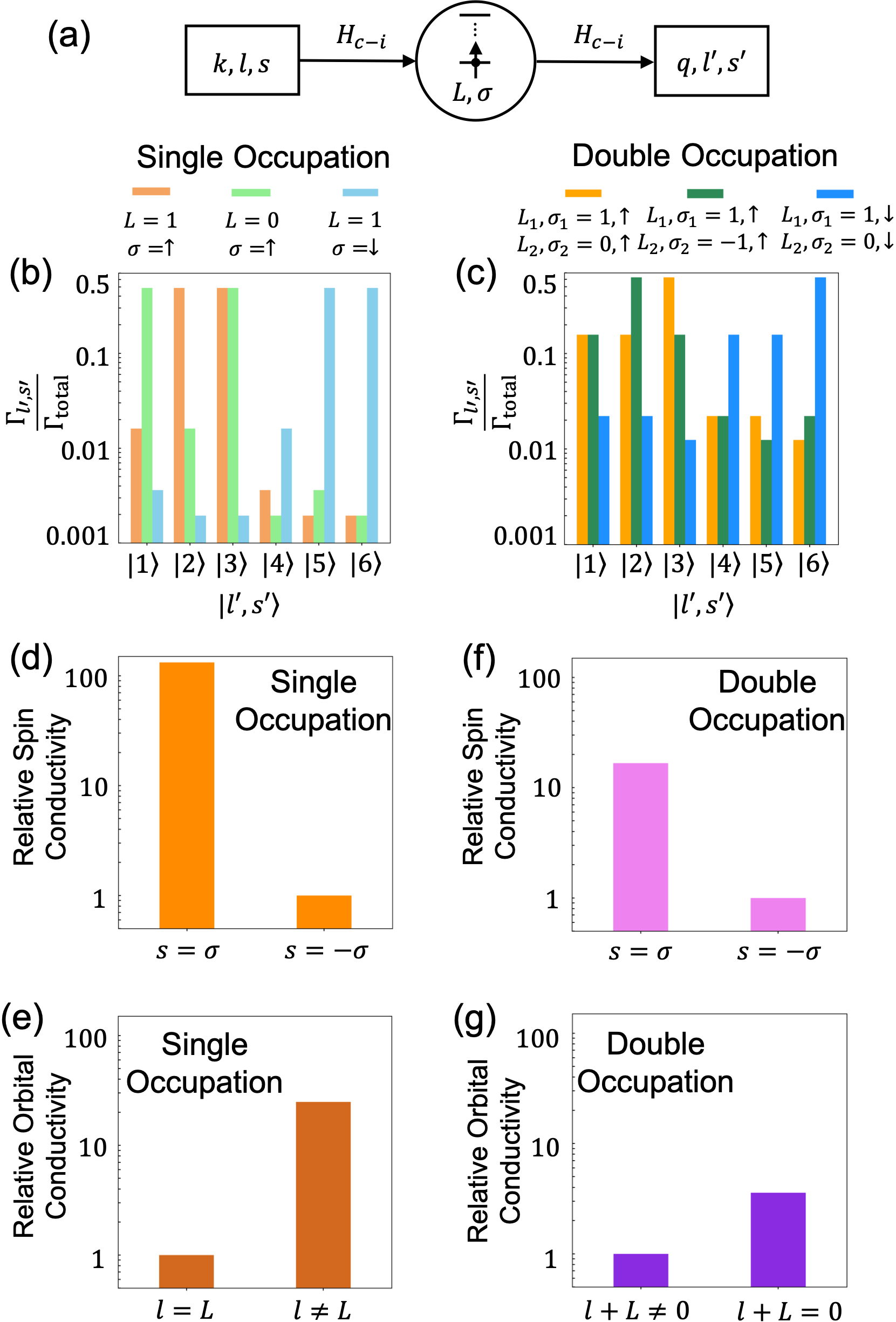}
\caption{(a) Illustration of the tunneling process. Conduction electrons in the left (right) lead has $k,\ell, s$ ($q,\ell', s'$) as their momenta, OAM, and spin. Electrons localized on the individual atoms within the junction have their orbital and spin angular momenta $L$ and $\sigma$. $H_{c-i}$ is the coupling Hamiltonian between the quantum dot and the leads. The overall SOFE is plotted in (b) and (c) for single occupation and double occupation of the localized states, respectively, which is the same for the initial and final states. The vertical axis indicates the tunneling rate of outgoing electrons scaled by the total tunneling rate, while the horizontal axis indicates the final states of the outgoing current, $\left| \ell', s' \right>  =  \left|1\right>, \left|2\right>, \left|3\right>, \left|4\right>, \left|5\right>, \left|6\right>  =  \left|1, \uparrow \right>, \left|0, \uparrow\right>, \left|-1, \uparrow\right>, \left|1, \downarrow\right>, \left|0, \downarrow\right>, \left|-1, \downarrow\right> $. Different colors correspond to different localized states in the junction. (d)-(g) plot the SOMR in terms of conductivity, where the horizontal axis indicates different polarizations of both the incoming current and the localized states. The conductivity in each figure is scaled by the smallest value of the conductivity. (d) and (f) show the spin conductivity, while (e) and (g) show the orbital conductivity.  (d) and (e) correspond to single occupation of the localized states, while (f) and (g) correspond to double occupation. 
}
\label{fig:filter_OMR}
\end{figure}

We consider a tunnel junction illustrated in Fig.~\ref{fig:filter_OMR}(a). The junction consists of many individual atoms hosting local states with spin and orbital degrees of freedom, which is weakly coupled to the left and right leads via hopping matrices. A conduction electron at the Fermi energy $\epsilon_F$ in the left lead can tunnel to an atom inside the junction with its energy elevated above $\epsilon_F$ due to Coulomb repulsion. As a result, the junction acts as an effective tunneling barrier that resists the current. The energy will be lowered as an electron hops from an individual atom in the junction into the right lead, thereby completing the tunneling process. An alternative tunneling process swaps the order of the above two sub-processes so the intermediate state has a lower energy. Overall, the prescribed spin and orbital states of the individual atoms inside the junction, assumed not to be altered by the tunneling, determines a set of possible intermediate states with different energies, hence affecting the SOFE and SOMR. We emphasize that the electron hopping among individual atoms is weak, so compared to the single-site tunneling, a tunneling event involving multiple localized sites is a higher order process which will be ignored. Once the transition rate is obtained for a single atom in the junction, the result can be straightforwardly generalized to an ensemble of individual atoms.

We denote the initial state by $\left| i \right>$ (before tunneling), the final state by $\left| f \right>$ (after tunneling), and the intermediate state by $\left| \lambda \right>$.  In our model, each state includes electrons in the left lead, the junction, and the right lead, which can be written as
\begin{align} \label{eq:quantum_state}
    &\left| L(N_{L}), C(N_{C}), R(N_{R}) \right>  \notag \\    
    &= \left( \bigotimes^{N_{L}}_{i=1} \left|k_{i}, \ell_{i},s_{i} \right> \right) \otimes \left( \bigotimes^{N_{C}}_{i=1} \left|L_{i},\sigma_{i} \right> \right)   \notag \\
    &\otimes \left( \bigotimes^{N_{R}}_{i=1} \left|q_{i}, \ell_{i}',s_{i}' \right> \right),  
\end{align}
where $N_{L}$, $N_{C}$, and $N_{R}$ are the number of electrons in the left lead, junction, and the right lead, respectively. The transition rate is given by~\cite{quantumtransport}
\begin{align}
    \gamma_{i\to f} = \frac{2\pi}{\hbar} \left| \sum_{\lambda} \frac{\left< f \right| H_{c-i} \left| \lambda \right> \left< \lambda \right| H_{c-i} \left|i \right> }{E_{i} - E_{\lambda}}  \right|^{2},
    \label{eq:t_rate_orig}
\end{align}
where $H_{c-i}$ is defined in Eq.~\ref{eq:hopping_H_k}. Energy conservation $E_i = E_f$ is automatically satisfied provided that the Fermi energies of the two leads are approximately equal. An elementary tunneling event corresponds to the transition
\begin{align}
 &\left| L(N_{L}), C(N_{C}), R(N_{R}) \right> \notag\\
 &\qquad\qquad\to \left| L'(N_{L}-1), C(N_{C}), R'(N_{R}+1) \right>
\end{align}
with the intermediate state being $\left| L'(N_{L}-1), C'(N_{C} + 1), R(N_{R}) \right>$ or $\left| L(N_{L}), C''(N_{C} - 1), R'(N_{R}+1) \right>$. Since SOFE and SOMR concern the influence of local occupations on transport, we require that the individual atomic states return to their original states after the tunneling. The tunneling rate calculated from Eq.~\eqref{eq:t_rate_orig} is $\gamma_{L \to R}$. This picture is analogous to the co-tunneling processes in Coulomb blockade. Because inelastic co-tunneling, involving incoherent intermediate states, typically dominates over elastic processes~\cite{quantumtransport}, we can approximate $|\sum_{\lambda} \cdots|^{2}$ as $\sum_{\lambda}|\cdots|^{2}$, i.e., summing probabilities rather than probability amplitudes. We further assume that a small voltage is added between the two leads to suppress backward tunneling ($\gamma_{R \to L}$) so that the tunneling current is entirely determined by $\gamma_{L \to R}$. In this section, $H_{c-i}$ uses the hopping matrix in Eq.~\eqref{eq:hopping_M_bcc} to ensure a close connection with realistic materials.

An alternative perspective to model the orbital-dependent transport is to consider these magnetic atoms embedded in the bulk of a conducting metal, so their influence on the SOFE and SOMR cannot be captured through individual tunneling processes. Instead, conduction electrons interact with localized states in a way that shifts the conduction band energies via second-order perturbation, which in turn modify both the density of states and the group velocities of electrons at the Fermi level. According to Boltzmann transport theory, such modifications lead to changes in conductivity. Because the energy shifts depend on both orbital and spin degrees of freedom, the resulting conductivities of conduction bands also acquire orbital and spin dependence. A more detailed discussion of bulk spin- and orbital-dependent conductivities is provided in the Supplementary Information~\ref{sec:bulk_transport}.

\subsection{Spin-Orbital Filter Effect}

The incoming electric current in the left lead is unpolarized, meaning that electrons with arbitrary $\ell$ and $s$ contribute equally. When passing through the junction, the interaction with localized electrons can polarize the outgoing current. We quantify this polarization by evaluating the outgoing tunneling rate of electrons with quantum numbers $\ell'$ and $s'$,
\begin{align} \label{eq:SOFE_rate}
    \Gamma^{\rm SOFE}_{\ell',s'} =\frac{k_B T\rho_{F}^{2} n_{C}}{2(2L_{\rm max}+1)}  \sum_{\ell,s} \gamma_{\ell,s \to \ell',s'},
\end{align}
where $\rho_{F}$ is the density of states at the Fermi level in both leads, $n_{C}$ is total number of individual atoms in the junction, $L_{\rm max}$ is the maximum possible orbital quantum number ($L_{\rm max}=1$ for the $p$-orbital), and $\gamma_{\ell,s \to \ell',s'}$ is the tunneling rate from a conduction band with $\ell$ and $s$ in the left lead to a conduction band with $\ell'$ and $s'$ in the right lead, as given in Eq.~\eqref{eq:t_rate_orig}. Here, $k_B$ is the Boltzmann constant, $T$ the temperature, and the factor $1/[2(2L+1)]$ reflects the uniform probability of incoming electrons due to their fully unpolarized distribution. We assume that the local atoms are fully polarized with uniform electron occupation, and the whole process takes place at sufficiently low temperature.

Here we set the model parameters as $t = 1$, $\epsilon_{F} = 0$, $\epsilon_{0} = -0.1$, $U = 4.5$, $U' = 1.5$, and $J = 0.5$ (all in units of eV)~\cite{lin2021origin}. A related case without Hund’s rules is discussed in the Supplementary Information~\ref{sec:no_hund_rule_comp}. Since the hopping amplitude $t$ appears only as a global prefactor in the final results, we set it to $1$ for simplicity. The calculated tunneling rates for different occupations are shown in Fig.~\ref{fig:filter_OMR}(b) and (c). In Fig.~\ref{fig:filter_OMR}(b), for single occupation, we choose $L,\sigma = 1, \uparrow; 0,\uparrow; 1, \downarrow$ as three representative cases, where $L,\sigma = 1, \uparrow$ is the reference, $L,\sigma = 0, \uparrow$ shows the impact of different orbital states with the same spin, and $L,\sigma = 1, \downarrow$ shows the impact of different spin states with the same orbital states. Similarly, we choose three representative cases in Fig.~\ref{fig:filter_OMR}(c) for double occupation. The outgoing current exhibits pronounced spin and orbital polarization. The spin aligns parallel to that of the localized states inside the junction, while the orbital degrees of freedom tend to complement those of the local states. By ``complement,” we mean that the OAM of the polarized current favors states different from those occupied in the junction. For example, in Fig.~\ref{fig:filter_OMR}(b), when the local atoms are singly occupied with $L = 1$, the outgoing current is dominated by an equal mixture of $L = 0$ and $L = -1$. Similarly, in Fig.~\ref{fig:filter_OMR}(c), when the local atoms are doubly occupied with $L = 1$ and $L = 0$, the outgoing current is dominated by $L = -1$.

The mechanism underlying the SOFE can be further understood by analyzing the spin and orbital dependence of the atoms inside the junction. According to Hund’s rules, among all possible intermediate occupations, those with localized electrons aligned in spin have the lowest energy. Consequently, tunneling processes involving such states have larger amplitudes. Moreover, since our model does not allow spin-flip hopping, spin is conserved throughout the tunneling process. These factors together ensure that conduction bands with the same spin orientation as the localized states exhibit higher tunneling rates. The orbital dependence is more subtle. Suppose that the local atoms host electrons with orbitals $\{L_{i}\}$, where $i=1,\dots,N_{C}$, and all spins are aligned. Different configurations have different energies due to Coulomb repulsion defined by Eq.~\eqref{eq:Hubbard_U}. One class of tunneling processes occurs when an electron from the left lead hops to the local site, creating an intermediate state with $N_{C}+1$ electrons. These states experience stronger Coulomb repulsion and thus lies at higher energies, leading to smaller tunneling amplitudes. Another class occurs when a local electron with orbital $L_{i}$ hops to the right lead first, reducing the occupation to $N_{C}-1$. These intermediate states are energetically lower than the initial ones and therefore contribute larger amplitudes. Within this more favorable class, the orbital dependence arises from the off-diagonal structure of the hopping matrix in Eq.~\eqref{eq:hopping_M_bcc}. Specifically, a local electron in orbital $L_{i}$ can hop into conduction bands with $\ell'\neq L_{i}$. If $\ell'$ does not belong to the occupied set $\{L_{i}\}$, the number of available processes is larger, which enhances the tunneling rate. Thus, orbitals complementary to the local occupation dominate the outgoing current. We emphasize that this analysis is strongly dependent on the detailed form of the hopping matrix $t^{a}_{\ell,L,\bm{d}}$. Different material systems, with different hopping structures, may yield qualitatively different orbital selection rules when analyzed in the same way.

In addition to the relative tunneling rates, the absolute magnitude of the tunneling rate, omitted in Fig.~\ref{fig:filter_OMR}, is reduced when the local states are doubly occupied compared to the singly occupied case. This reduction arises because the involved intermediate states for a doubly occupied atom lie at higher energies, which suppresses the transition amplitude and thus lowers the tunneling rate. We have also tested alternative parameter regimes in which Hund’s rules almost break down such that Eq.~\eqref{eq:impurity_H} becomes nearly independent of spin and orbital indices. The corresponding results are presented in the Supplementary Information~\ref{sec:no_hund_rule_comp}.

\subsection{Spin Orbital Magnetoresistivity}

For SOMR, the incoming electric current in the left lead is spin or orbital polarized in priori, while the localized states of the magnetic atoms are still assumed to be the same during the process. The conductivity therefore depends on the spin and orbital angular momenta of both the incoming electrons and those occupying the localized states. We characterize SOMR by the tunneling rate,
\begin{align} \label{eq:SOMC_rate}
    \Gamma^{\rm SOMR}_{\ell,s} =k_B T\rho_{F}^{2} n_{C}  \sum_{\ell',s'} \gamma_{\ell,s \to \ell',s'}.
\end{align}
where the notation follows that in Eq.~\eqref{eq:SOFE_rate}. The key distinction between the two cases is in the summation. For SOMR the sum runs over outgoing states in the right lead, whereas for SOFE the sum runs over incoming states in the left lead.

Similar to the previous section, we discuss the case of single and double occupations separately. To make our results more intuitive, we separately calculate the orbital-only and spin-only tunneling rates by further summing over $s$ ($\ell$) of the incoming electrons and $\sigma$ ($L$) of the localized state. This corresponds to assuming a fully orbital- (spin-) polarized and fully spin- (orbital-) unpolarized incoming current and the configuration of localized states. The results are shown in Fig.~\ref{fig:filter_OMR}(d)–(g). In Fig.~\ref{fig:filter_OMR}(g), $\ell+L$ denotes the combined orbital quantum number of the incoming current and localized state. For example, if the incoming current is fully polarized with $\ell=-1$ while the local states are polarized with $L=1$, then $\ell+L=0$. The results reveal that spin MR is significantly larger when the spin of the incoming conduction electron matches that of the localized states. In contrast, orbital MR is strongly suppressed if the orbital polarization of the incoming current overlaps with the occupied orbitals in the local states. The contrast between high and low MR is therefore more pronounced for spin than for orbitals. For example, for double occupation, the relative spin MR is $\sim 1600\%$ while the relative orbital MR is $\sim 360\%$. Moreover, this contrast diminishes as the localized states host more electrons. Overall, the behavior of SOMR is consistent with the SOFE trends discussed earlier, and the mechanism analyzed in the previous section can also explain results in this section. We have also tested parameter values for which Hund’s rules are nearly broken; the corresponding SOMR results are provided in the Supplementary Information~\ref{sec:no_hund_rule_comp}.

To conclude this section, we note that localized states with triple occupation do not exhibit orbital-dependent SOFE or SOMR. This is because the ground state consists of three electrons occupying $L = 1, 0, -1$ with aligned spins. Such configuration is symmetric with respect to all orbital degrees of freedom, hence suppressing the orbital polarization. This result is consistent with the vanishing orbital exchange regarding the exchange interaction of triple occupation. Consequently, both SOFE and SOMR become purely spin dependent. The corresponding spin-dependence for the triple occupation case is presented in the Supplementary Information~\ref{sec:triple_occupy}.

\section{Switching by Orbital Torque}

As a reciprocal effect of SOFE and SOMR, a spin- or orbital-polarized current can exert torques on the magnetic atoms and switch the localized states. Their reciprocal relation is analogous to, and an important generalization of, that between spin-only MR and spin-torque-induced switching. In this section, we first establish a formalism to describe the switching dynamics of localized states by an orbital polarized current with various spin polarization. We then demonstrate the orbital-current-induced switching in the simple $p$-orbital states.

\subsection{Model}

We consider the same device geometry as illustrated in Fig.~\ref{fig:filter_OMR}(a). The physical process of switching is similar in spirit to SOFE and SOMR, but with one key difference: here we allow the localized states of the individual atoms in the junction to change under the action of orbital torques during the tunneling process. Using the notation of Eq.~\eqref{eq:quantum_state}, each tunneling event corresponds to a transition
$\left| L, C, R \right> \to \left| L', C', R' \right>$, whereas in SOFE and SOMR the local configuration remains unchanged, i.e., $\left| L, C, R \right> \to \left| L', C, R'\right>$. The transition rate $\gamma_{\ell,s;\alpha \to \ell',s';\alpha'}$ denotes the process in which an incoming electron with $\ell$ and $s$ in the left lead tunnels towards a conduction band with $\ell'$ and $s'$ in the right lead, while the occupations of localized states switch from $\alpha$ to $\alpha'$ (here $\alpha$ labels a local configuration like $L_{1}=1,\sigma_{1}=\uparrow, L_{2} = 0,\sigma_{2}=\uparrow$). This tunneling rate is evaluated using Eq.~\eqref{eq:t_rate_orig}. Because we now focus on the switching of occupations of the localized states rather than the outgoing electrons, we need to sum over all possible final itinerant states in the right lead. This yields an effective transition rate between different configurations,
\begin{align} \label{eq:effect_rate}
    \Gamma_{\alpha \to \alpha'} =k_B T\rho_{F}^{2} n_{C} \sum_{\ell',s'} \gamma_{\ell,s; \alpha\to \ell', s'; \alpha'},
\end{align}
which is similar in form to Eq.~\eqref{eq:SOMC_rate} but the summation runs over a different range. In fact, when $\alpha = \alpha'$, Eq.~\eqref{eq:effect_rate} reduces to Eq.~\eqref{eq:SOMC_rate}. It should be noted that Eq.~\eqref{eq:effect_rate} can be readily generalized to the case of partially polarized or unpolarized incoming currents. In that case, an additional summation over the incoming quantum numbers $\ell,s$ is required.

With the local atoms uniformly polarized in the initial state $\alpha$, tunneling events can switch some atoms to a new configuration $\alpha'$. These atoms can subsequently transition to $\alpha''$ or revert back to $\alpha$, followed by further transitions. Such sequential evolution among different states is in fact a stochastic process, which calls for a modification of the formalism established in previous sections. Let $P_{\alpha}$ denote the fraction of atoms in configuration $\alpha$, whose evolution dynamics is governed by
\begin{align} \label{eq:P_eq_orig}
    \frac{dP_{\alpha}}{dt} = \sum_{\alpha'} P_{\alpha'} (1-P_{\alpha}) \Gamma_{\alpha' \to \alpha},
\end{align}
where the term $1-P_{\alpha}$ accounts for the proportion of the unoccupied states, reflecting Pauli's exclusion principle. A similar factor also appears in the expressions for tunneling currents for the same reason~\cite{quantumtransport}. Equation~\eqref{eq:P_eq_orig} can be recast into a nonlinear Markov process by introducing the generator matrix with elements:
\begin{align} \label{eq:Q_def}
    Q_{\alpha' \to \alpha}=
    \begin{cases}
    (1-P_{\alpha}) \Gamma_{\alpha' \to \alpha}, \text{if } \alpha \neq \alpha' \\
       -\sum_{\alpha\neq \alpha'}Q_{\alpha' \to \alpha}, \text{if } \alpha = \alpha',
    \end{cases}
\end{align}
which are subject to the probability normalization of Markov-chain generators~\cite{nonlinearMarkov}. With this definition, Eq.~\eqref{eq:P_eq_orig} becomes
\begin{equation} \label{eq:P_eq}
    \frac{dP_{\alpha}}{dt} = \sum_{\alpha'} P_{\alpha'} Q_{\alpha' \to \alpha},
\end{equation}
which is a set of nonlinear equations since $Q_{\alpha' \to \alpha}$ depends on $P_\alpha$. Analytical solutions to Eq.~\eqref{eq:P_eq} are in general intractable, so we proceed with numerical solutions.

\subsection{Switching Behavior}

\begin{figure}[!htbp]
\centering
\includegraphics[width=1\linewidth]{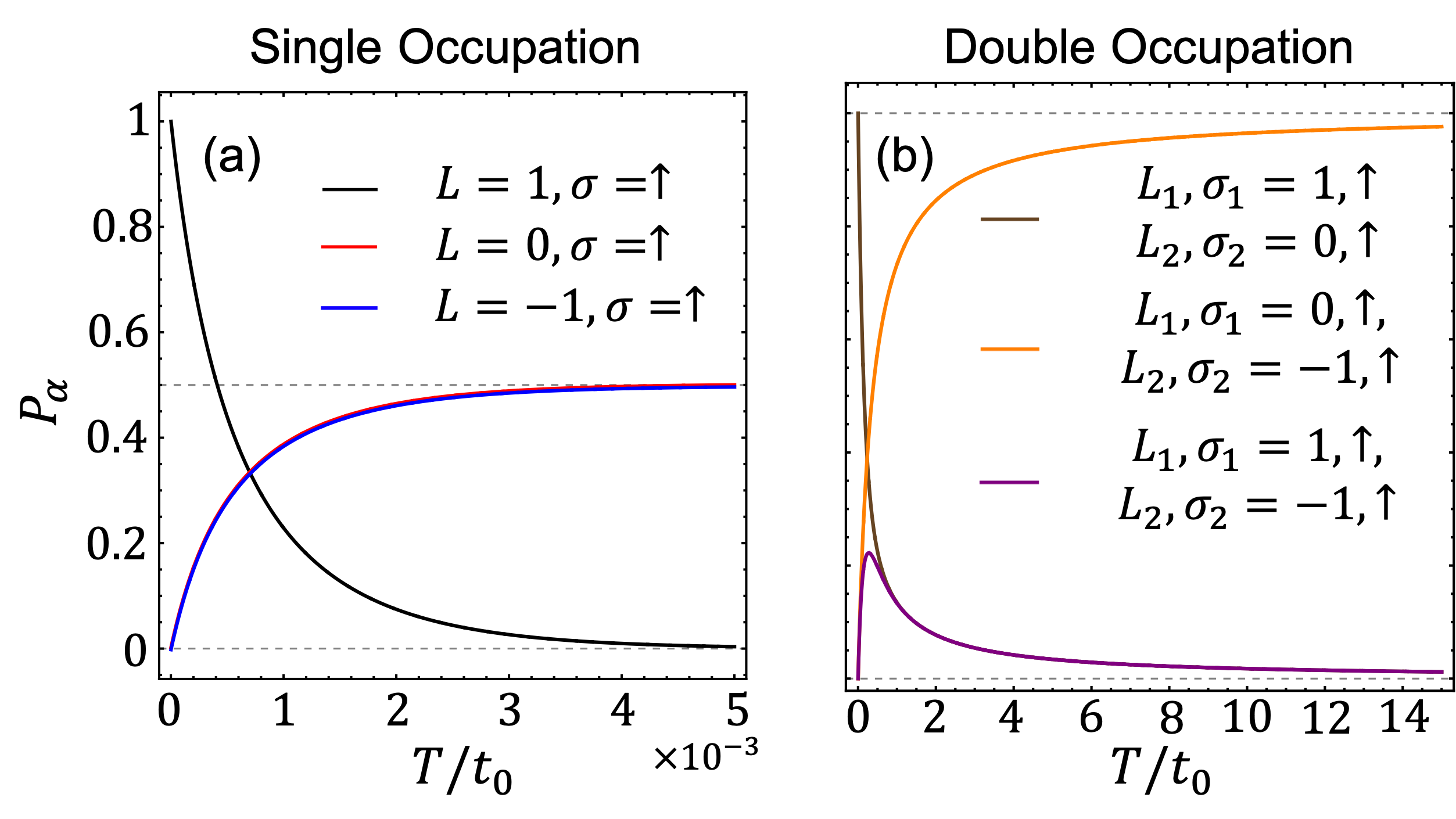}
\caption{Time evolution of $P_{\alpha}$ in the presence of a fully polarized incoming current in the left lead with $\ell = 1, s=\uparrow$. Localized states are initially fully polarized to (a) $L=1, \sigma = \uparrow$ and (b) $L_{1}=1, \sigma_{1} = \uparrow, L_{2}=0, \sigma_{2} = \uparrow$. $t_0$ is the unit of time $T$, and $t_0 = \hbar / (2\pi k_B T \rho_F^2 n_{C} t^{2})$, where the definitions of symbols are provided in the main text. Curves corresponding to $L = 0, \sigma = \uparrow$ and $L = -1, \sigma = \uparrow$ in (a) are slightly shifted for better clarity.
}
\label{fig:switching}
\end{figure}

We perform numerical simulations, and the time evolution of $P_{\alpha}$ for specific initial configurations is shown in Fig.~\ref{fig:switching}. The incoming current is fully polarized with $\ell = 1$ and $s = \uparrow$. In both Fig.~\ref{fig:switching}(a) and (b), the local atoms are initially fully polarized. The proportion of the initial configuration decreases over time until eventually vanishes, while other configurations grow with time. In Fig.~\ref{fig:switching}(a), the final states are shared equally between $L = 0, \sigma = \uparrow$ and $L = -1, \sigma = \uparrow$, each reaching $P = 0.5$. Their growth proceeds in the same pace, while all other configurations remain absent throughout the evolution. By contrast, in Fig.~\ref{fig:switching}(b), the dynamics is more intricate. Initially, both configurations of double occupations $L_{1} = 0, \sigma_{1} = \uparrow, L_{2} = -1, \sigma_{2} = \uparrow $ and $L_{1} = 1, \sigma_{1} = \uparrow, L_{2} = -1, \sigma_{2} = \uparrow $ increase in proportion. However, at later times, $L_{1} = 1, \sigma_{1} = \uparrow, L_{2} = -1, \sigma_{2} = \uparrow $ reverses its growth and declines, sending the system into $L_{1} = 0, \sigma_{1} = \uparrow, L_{2} = -1, \sigma_{2} = \uparrow $ as time further increases. This reversal can be understood from the microscopic hopping processes. According to Eq.~\eqref{eq:hopping_M_bcc}, incoming electrons can only hop into the empty $L = 0$ and $L = -1$ orbitals without spin flip. Thus, $L_{1} = 0, \sigma_{1} = \uparrow, L_{2} = -1, \sigma_{2} = \uparrow$ cannot absorb an additional electron to form a triply occupied state. Moreover, after emitting an electron into the right lead, the intermediate configuration is either $L = 0, \sigma = \uparrow$ or $L = -1, \sigma = \uparrow$, and subsequent absorption always restores $L_{1} = 0, \sigma_{1} = \uparrow, L_{2} = -1, \sigma_{2} = \uparrow$. This explains why its proportion eventually saturates to $1$. By contrast, the initial configuration can first transition to $L_{1} = 1, \sigma_{1} = \uparrow, L_{2} = -1, \sigma_{2} = \uparrow$, but this configuration can then further switch irreversibly to $L_{1} = 0, \sigma_{1} = \uparrow, L_{2} = -1, \sigma_{2} = \uparrow$, causing its eventual decline. Another distinction between Fig.~\ref{fig:switching}(a) and (b) lies in the switching timescale, which turns out to be much shorter for the single-occupation case. This property is attributed to the lower energy of intermediate states, which yields larger transition rates $\gamma_{i \to f}$ in Eq.~\eqref{eq:t_rate_orig} and thus faster dynamics.

We have tested different initial localized configurations (of magnetic atoms) and found that their final configurations are independent of the initial configurations; only the polarization of the incoming current matters. This observation is consistent with the spin-only case, where the equilibrium local spin configurations only depends on the spin polarization of the incoming current regardless of the initial configuration of localized spins. The equilibrium configurations are summarized in Table~\ref{tab:switching_final_state}. A clear pattern emerges: local atoms evolve into configurations that do not share orbital degrees of freedom with the incoming itinerant electrons. For single occupations, this means the equilibrium configuration satisfies $L \neq \ell$ and $\sigma = s$, with equal proportions of $P_{\alpha}$. For double occupations, only configurations of the form $L_{i} \neq \ell, \sigma_{i} = s$ survive. This conclusion remains robust against different parameter choices. Note that for double occupations, when the localized spins are opposite to those of the incoming electrons (e.g., $\sigma_i = -s$), these configurations cannot be switched and thus remain unchanged throughout the process. This is because the lowest-order hopping can flip only one local spin. According to Eq.~\eqref{eq:Hubbard_U}, the resulting intermediate states lie at higher energy than the initial ones. Energy conservation therefore forbids such switching. For triple occupations, with similar reasoning, there is no switching of local configurations either, since the ground state configuration of triple occupations is three electrons with all spin up or down. If we want to switch one configuration to another, all spins need to be flipped, which is a very high order process.

\begin{table}
\caption{\label{tab:switching_final_state}%
Equilibrium configurations in the presence of a fully polarized incoming current in the left lead. The first column indicates the orbital and spins of the incoming current. The second and third columns indicate the local states are singly or doubly occupied. 
}
\begin{ruledtabular}
\begin{tabular}{ccc}
\textrm{$\ell, s$}&
\textrm{Single Occupation $L, \sigma$}&
\textrm{Double Occupation $L_{i}, \sigma_i$\footnote{If the local spins are different from the incoming current, these configurations remain unchanged.}}\\
\colrule
\multirow{2}{*}{$1, \uparrow$} & $50\%\, 0, \uparrow$  & $0, \uparrow,$ \\
& $50\%\, -1,\uparrow$ & $ -1, \uparrow $ \\
\multirow{2}{*}{$0, \uparrow$} & $50\%\, 1, \uparrow$ & $ 1, \uparrow,$ \\
 & $50\%\, -1, \uparrow$  & $  -1,\uparrow$ \\
\multirow{2}{*}{$-1, \uparrow$} & $50\%\, 1, \uparrow$ & $ 1,\uparrow,$\\
& $50\%\, 0, \uparrow$ & $0, \uparrow$ \\
\multirow{2}{*}{$1, \downarrow$} & $50\%\, 0, \downarrow$ & $ 0, \downarrow,$ \\
& $50\%\, -1, \downarrow$ & $ -1, \downarrow $ \\
\multirow{2}{*}{$0, \downarrow$} & $50\%\, 1,\downarrow$ & $ 1, \downarrow,$ \\
& $50\%\, -1, \downarrow$ & $ -1, \downarrow $ \\
\multirow{2}{*}{$-1, \downarrow$} & $50\%\, 1, \downarrow$ & $ 1, \downarrow,$\\
& $50\%\, 0, \downarrow$ & $ 0, \downarrow $ \\
\end{tabular}
\end{ruledtabular}
\end{table}

\section{Conclusion and Outlook}

In this work, we have established a proof-of-concept orbital exchange interaction from a microscopic perspective by generalizing the Anderson–Kondo model to include orbital degrees of freedom. The result is qualitatively different from, hence cannot be expressed similarly as, the conventional $s$–$d$ exchange. In particular, the orbital exchange strongly depends on the occupation of localized states of the magnetic atoms and their spatial structure. As a simple demonstration, we considered a diagonal hopping matrix to derive the orbital exchange and classified the results into three categories: spin-only, orbital-only, and spin–orbital exchange. We then studied the spin- and orbital-dependent transport based on a simple tunneling junction device, where we predicted the spin-orbital filtering and spin-orbital MR, with pronounced dependence on the occupation of the localized orbitals. Reciprocally, we have explored the orbital switching induced by orbital torques, where both spin and orbital degrees of freedom can be altered by an orbital polarized current. To capture the cumulative consequence of multi-step orbital switching, we formulated a nonlinear Markov model compatible with Pauli's exclusion principle. Numerical simulations reveal that the equilibrium configurations are independent of the initial configurations and are determined solely by the polarization of the incoming current.

In analyzing the orbital-dependent transport, we calculated the hopping matrix in a body-centered cubic lattice, which renders off-diagonal components in the hopping matrix. Our formalism can be readily extended to study more realistic material circumstances, calling for first-principle calculations to determine the hopping matrix alongside other parameters. We anticipate that our study will inspire a series of future investigations based on real materials. With the physical picture established in this work, it is essential to revisit several conclusions reached in previous studies that were based on the problematic form of orbital exchange interaction, such as orbital pumping and time-domain orbital dynamics driven by orbital torques. Another important direction in future studies is to go beyond the intra-atomic contribution to the OAM considered in this work.

\begin{acknowledgments}
The authors are grateful to Kyung-Jin Lee and Mark Stiles for helpful discussions. 
This work is supported by the National Science Foundation under Award No. DMR-2339315.
\end{acknowledgments}

\appendix
\section{Bulk Picture of Orbital-Dependent Transport} \label{sec:bulk_transport}

In the main text, we investigate the spin–orbital filter effect (SOFE) and spin–orbital magnetoresistivity (SOMR) in a tunneling junction, where electron transport is mediated by localized states on individual atoms. An alternative scenario arises when individual atoms are embedded directly in the bulk of a metal, in which case no tunneling junction is present. Under the bulk picture, individual magnetic atoms shift the conduction-band energies, thereby leading to orbital-dependent transport.

We first consider how individual atoms modify the dispersion of the conduction band. Since the hopping amplitude $t_{\ell,L}$ is typically much smaller than both the Fermi energy and the local state energies, the coupling Hamiltonian $H_{c-i}$ in Eq.~\eqref{eq:Hamiltonian} can be treated as a perturbation to $H_c + H_i$. Because $H_{c-i}$ is off-diagonal, the first-order perturbation vanishes, and we must apply second-order perturbation theory. This procedure is equivalent to applying first-order perturbation theory to the effective Hamiltonian obtained through the second-order Schrieffer–Wolff transformation in Sec.~\ref{sec:effective_orbital_exchange}. The second-order energy correction to a system in state $\left| \mu \right>$ is
\begin{equation} \label{eq:energy_perturb}
    \Delta E_{\mu} = \sum_{\lambda \neq \mu} \frac{\left| \left< \lambda \right| \Delta H \left| \mu \right> \right|^{2}}{E_{\mu} - E_{\lambda}}.
\end{equation}
Although this correction applies to the full system, when focusing on conduction-electron transport it is natural and consistent with the spin-only case to interpret $\Delta E_{\mu}$ as a shift of the conduction-band dispersions. These shifts, being spin- and orbital-dependent, directly modify the density of states and group velocities of electrons at the Fermi level, and thus lead to orbital-dependent conductivity in bulk materials.

Within the Boltzmann transport framework, the current in each conduction band is given by
\begin{subequations} \label{eq:bulk_current_orig}
\begin{align}
    j_{\ell,s} &= e \int \frac{d^{3}k}{(2\pi)^3} \frac{\partial \epsilon_{k,\ell,s}}{\hbar \partial k} \delta f_{k,\ell,s},\\
    \delta f_{k,\ell,s} &= e\frac{\partial f_{\ell,s}(\epsilon)}{\partial \epsilon} v_{k,\ell,s} \tau_{k,\ell,s} E
\end{align}
\end{subequations}
where $\delta f_{k,\ell,s}$ is the non-equilibrium correction to the distribution function induced by an electric field $E$, $v_{k,\ell,s}$ is the group velocity, and $\tau_{k,\ell,s}$ the relaxation time. At low temperatures, $\partial f/\partial \epsilon$ can be approximated as a delta function at the Fermi level. Assuming isotropic bands in the $k$-space, the current simplifies to
\begin{equation} \label{eq:bulk_current_sim}
    j_{\ell,s} \propto e^{2} D_{\ell,s}(\epsilon_F) v^{2}_{\ell,s}(\epsilon_F) \tau E,
\end{equation}
where $D_{\ell,s}$ is the density of states and $\tau$ has been approximated as a constant relaxation time, and $v^{2}_{l,s}$ is obtained from averaging $v^{2}_{k,l,s}$. Because the corrections induced by localized states effectively shift the conduction-band dispersions, the energy change $\Delta E_{\ell,s}$ can be interpreted as the relative Fermi energy measured from the shifted band edge, $\Delta \epsilon_F = - \Delta E_{\ell,s}$. Consequently, the correction to the current is
\begin{align} \label{eq:bulk_current_expansion}
    \Delta j_{\ell,s} =& -e^{2} \tau E \left[ \frac{\partial D_{\ell,s}(\epsilon_F)}{\partial \epsilon} v^{2}_{\ell,s}(\epsilon_F) \right. \notag \\
    &\qquad\left. + D_{\ell,s}(\epsilon_F) \frac{\partial v^{2}_{\ell,s}(\epsilon_F)}{\partial \epsilon}  \right] \Delta E_{\ell,s},
\end{align}
which shows explicitly how the band shifts induced by individual magnetic atoms modify the conductivity. Because $\Delta E_{\ell,s}$ is orbital- and spin-dependent, the corrections to the conductivity inherit the same dependence, thereby giving rise to orbital-dependent transport in bulk systems.

To estimate the orbital-dependent current and conductivity, we adopt a simplified model of the conduction band. Assuming a quadratic dispersion, $\epsilon_{k,\ell,s} = \hbar^2 k^2 / (2m_e)$ with effective electron mass $m_e$, the current correction becomes
\begin{equation} \label{eq:bulk_current_q_model}
    \Delta j_{\ell,s} \propto- e^2 \tau E \frac{3}{4 \pi^2} \left(\frac{2m_e}{\hbar^2}\right)^{3/2} \frac{\sqrt{\epsilon_F}}{m_e} \Delta E_{\ell,s},
\end{equation}
where $\epsilon_F$ should be interpreted as the energy difference between the conduction-band edge and the Fermi level, which is different from the Fermi energy used to calculate energy correction. Since we are interested in the relative current corrections, common multiplicative factors are omitted in Eqs.~\eqref{eq:bulk_current_sim} and \eqref{eq:bulk_current_q_model}, as they cancel upon normalization. %

\begin{figure}[t]
\centering
\includegraphics[width=1\linewidth]{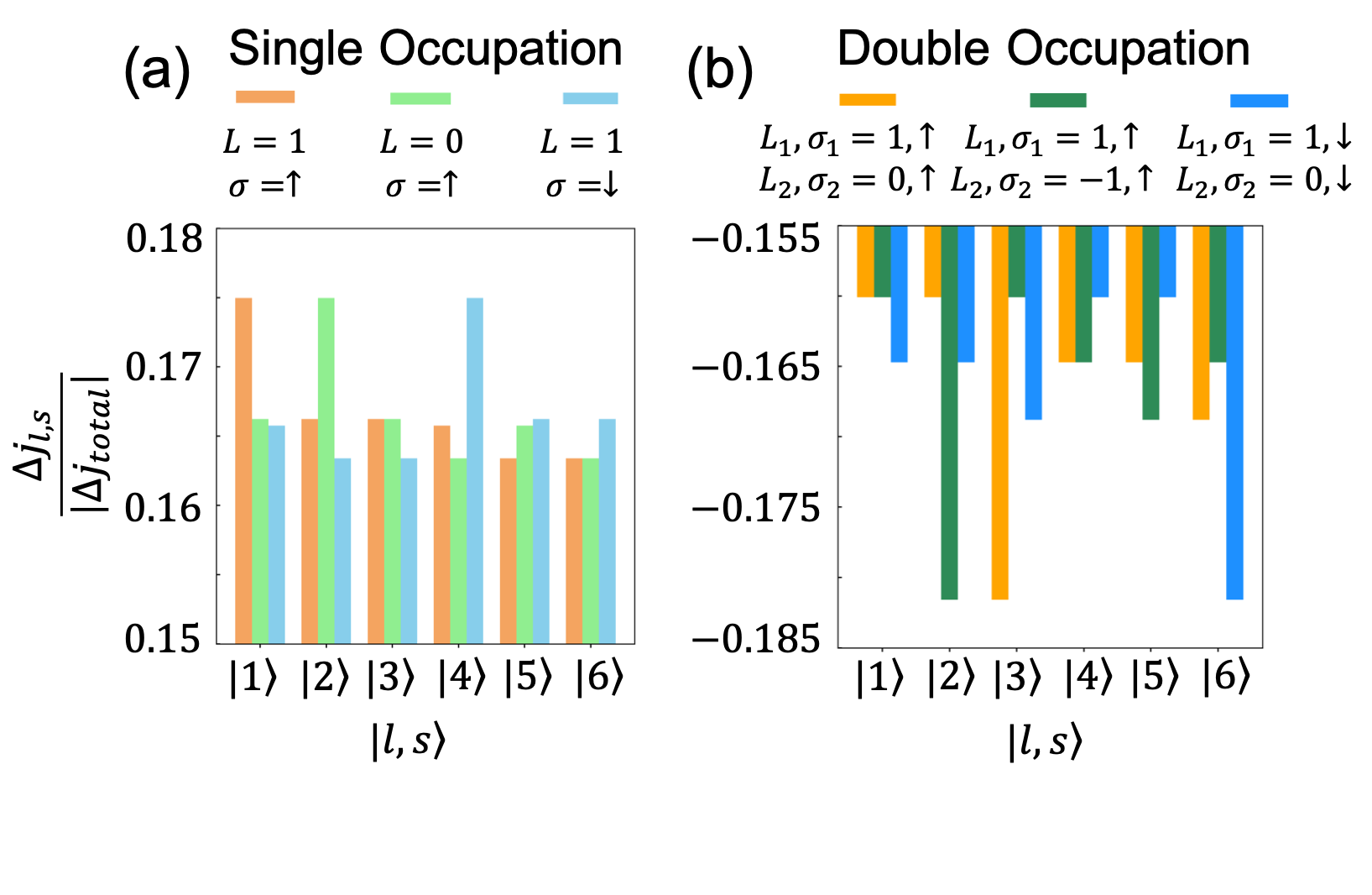}
\caption{Corrections to orbital- and spin-dependent current induced by the individual magnetic atoms in the bulk. (a) and (b) correspond to the single occupation and double occupation, respectively. The vertical axis shows the normalized current correction, $\Delta j_{\ell,s}/|\Delta j_{\mathrm{total}}|$, where $\Delta j_{\rm total} = \sum_{\ell, s}\Delta j_{\ell, s}$. The horizontal axis indicates the orbital and spin quantum numbers of conduction electrons, $\left| \ell, s \right>  =  \left|1\right>, \left|2\right>, \left|3\right>, \left|4\right>, \left|5\right>, \left|6\right>  =  \left|1, \uparrow \right>, \left|0, \uparrow\right>, \left|-1, \uparrow\right>, \left|1, \downarrow\right>, \left|0, \downarrow\right>, \left|-1, \downarrow\right> $. Different colors denote different configurations of the localized states.
}
\label{fig:per_eng}
\end{figure}

The orbital-dependent currents obtained from Eq.~\eqref{eq:bulk_current_q_model} are shown in Fig.~\ref{fig:per_eng}. A comparison with the tunneling case depicted in Fig.~\ref{fig:filter_OMR}(b) and (c) highlights both similarities and differences: while strong orbital dependence persists, its detailed form in the bulk differs from that in tunneling junctions. Notably, for single occupation [Fig.~\ref{fig:per_eng}(a)], $\Delta E_{\ell,s}<0$ and thus $\Delta j_{\ell,s}>0$, whereas for double occupation [Fig.~\ref{fig:per_eng}(b)], $\Delta E_{\ell,s}>0$ and consequently $\Delta j_{\ell,s}<0$.

\begin{figure}[t]
\centering
\includegraphics[width=1\linewidth]{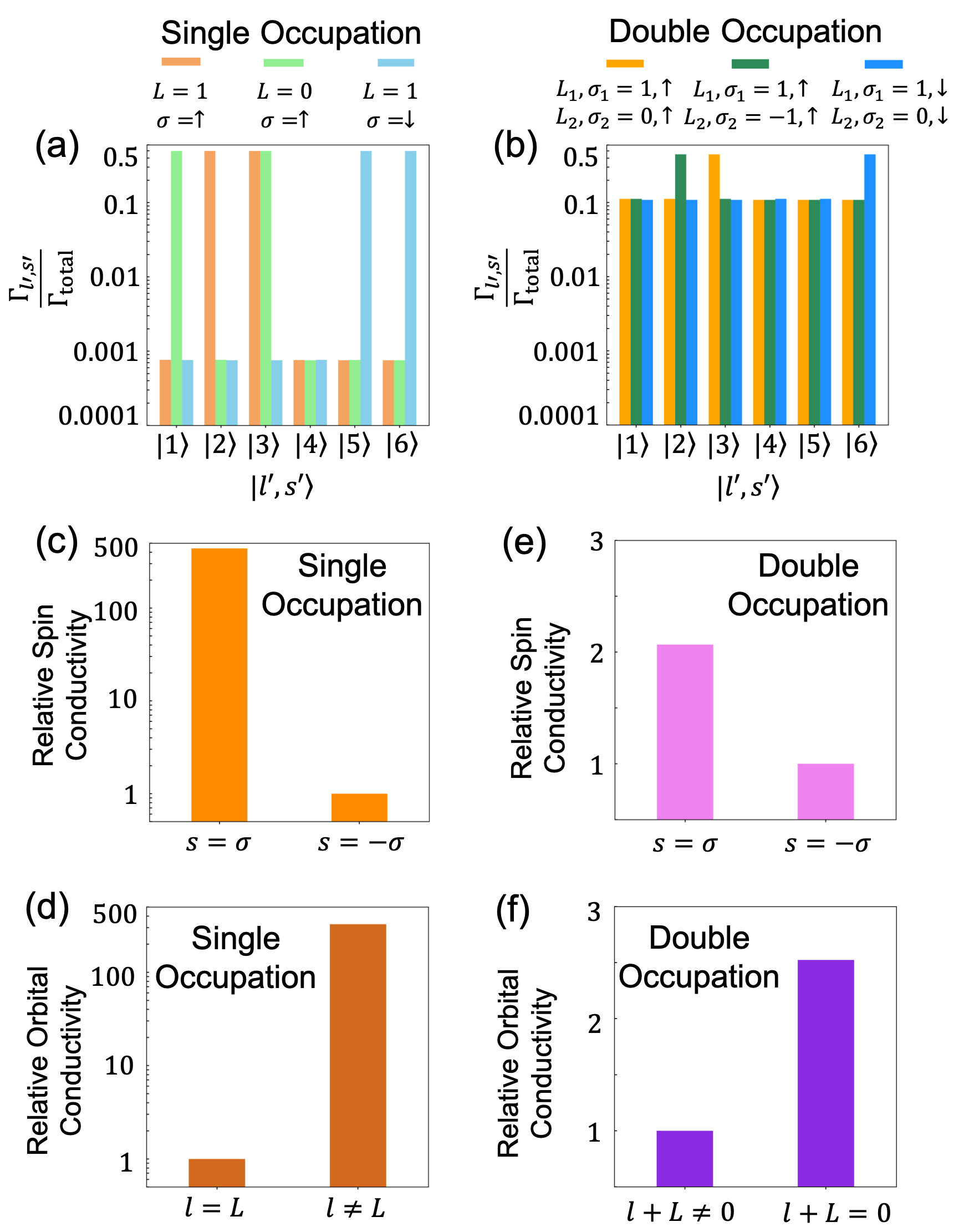}
\caption{SOFE and SOMR with strongly suppressed Hund's splitting. (a) and (b) show the SOFE for single and double occupations of the localized states, respectively. The vertical axis shows the outgoing tunneling rate normalized by the total rate, while the horizontal axis labels the orbital and spin quantum numbers of the outgoing electrons, following the convention in Fig.~\ref{fig:filter_OMR}. Different colors indicate different configurations of the localized states. (c)–(f) show the SOMR expressed in terms of conductivity, scaled by the minimum conductivity in each panel. (c) and (e) show the spin-dependent conductivity, while (d) and (f) show the orbital-dependent conductivity. (c) and (d) correspond to single occupation, whereas (e) and (f) correspond to double occupation of the localized states.
}
\label{fig:filter_OMR_no_Hund}
\end{figure} 

\begin{figure}[t]
\centering
\includegraphics[width=0.65\linewidth]{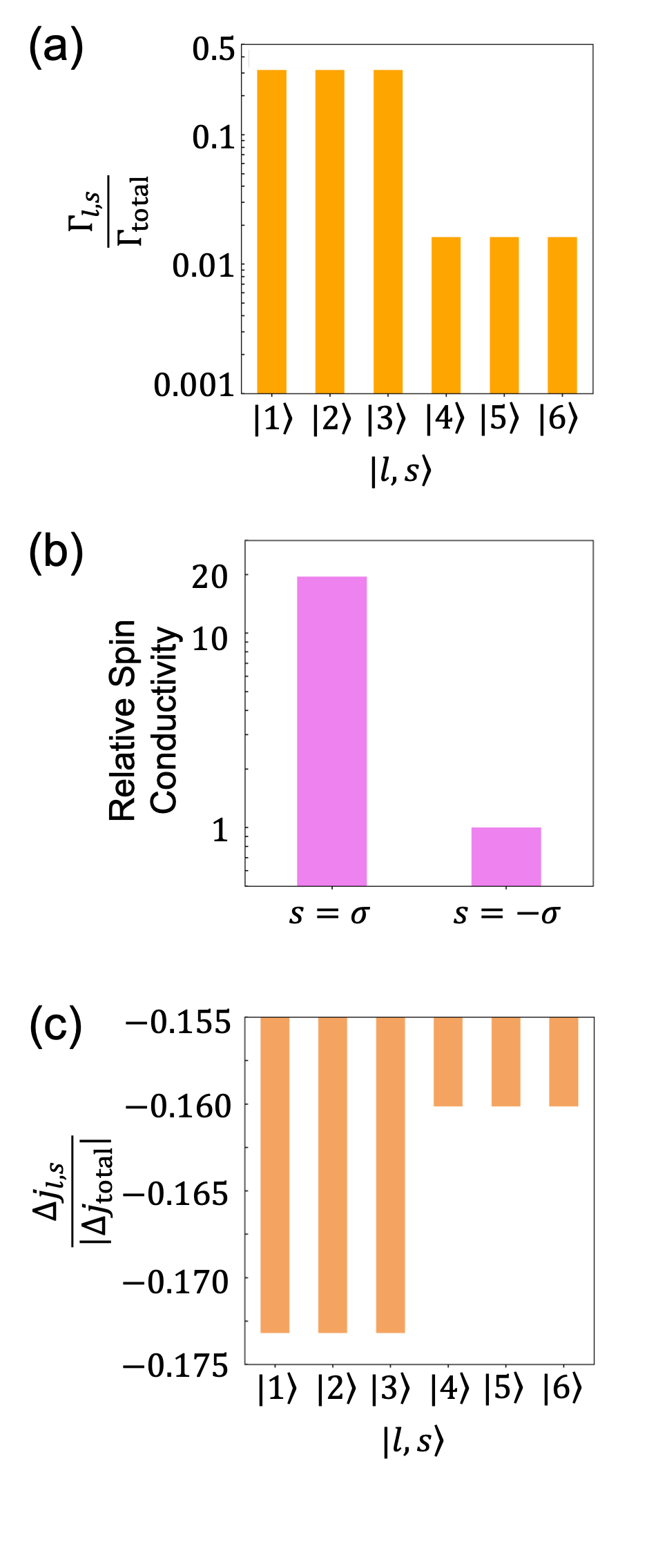}
\caption{ Orbital transport with triple occupation of the localized states. (a) shows the SOFE, where the vertical axis represents the tunneling rate of outgoing electrons normalized by the total rate, after the electrons pass through the junction. (b) shows the spin-dependent conductivity, where the horizontal axis indicates relative polarizations of both incoming current and localized states. The conductivity is scaled by the smallest value of conductivity. (c) shows the normalized corrections to orbital- and spin-dependent current induced by the individual atoms in a bulk material, where $\Delta j_{\rm total} = \sum_{\ell, s}\Delta j_{\ell, s}$. The horizontal axes in (a) and (c) indicate the orbital and spin quantum numbers of the outgoing and bulk conduction electrons, respectively, with six states $\left| \ell, s \right>$ corresponding to $ \left|1\right>, \left|2\right>, \left|3\right>, \left|4\right>, \left|5\right>, \left|6\right> = \left|1, \uparrow \right>, \left|0, \uparrow\right>, \left|-1, \uparrow\right>, \left|1, \downarrow\right>, \left|0, \downarrow\right>, \left|-1, \downarrow\right>$. The localized state has triple occupation of $L_{1} = 1, \sigma_{1} = \uparrow, L_{2} = 0, \sigma_{2} = \uparrow, L_{3} = -1, \sigma_{3} = \uparrow$.
}
\label{fig:triple}
\end{figure}

\section{Influence of Hund's-Rule Splitting on Orbital Transport} \label{sec:no_hund_rule_comp}

In the main text, we incorporate Hund’s rules through Eq.~\eqref{eq:Hubbard_U}. Because Hund’s rules split the energies of different intermediate states during both conduction-electron transport and switching of localized states, they can strongly affect orbital transport. To assess the impact of Hund's rules, especially when they are strongly suppressed (but none-zero to avoid orbital quenching), here we compare the results obtained in the main text with those calculated in the limit of nearly vanishing Hund's splitting. 
We choose parameters $t = 1$, $\epsilon_{F} = 0$, $\epsilon_{0} = -0.1$, $U = 4.5$, $U' = 4.48$, and $J = 0.01$ (in eV) to represent a system with suppressed Hund's splitting, while retaining the same hopping matrix as in the main text. The results are shown in Fig.~\ref{fig:filter_OMR_no_Hund}. Comparing Fig.~\ref{fig:filter_OMR_no_Hund} with Fig.~\ref{fig:filter_OMR}, we find that strong orbital dependence persists. However, in the limit of suppressed Hund's splitting, only two bands acquire significantly enhanced tunneling rates, while the rest are nearly indistinguishable. This arises because the energy differences between intermediate states become negligible, leaving the hopping matrix and Pauli's exclusion principle as the dominant factors. As a result, the spin and orbital MR are nearly equal. By contrast, when Hund’se splittings are not suppressed, the spin MR exhibits a stronger contrast than the orbital MR.

\section{Orbital Transport with Triple Occupation} \label{sec:triple_occupy}

We examine orbital transport with triple occupation using the same parameters as in the main text, and show the results in Fig.~\ref{fig:triple}. For clarity, we only display the spin-up case; the spin-down result follows directly by exchanging the spin-up and spin-down conduction bands. From Fig.~\ref{fig:triple}, we observe that when the localized states are triply occupied, the transport exhibits spin dependence but no orbital dependence. In addition, no switching occurs in this case. The underlying reason is that we restrict ourselves to the lowest-order switching processes, in which only a single electron in the localized states is changed. Within the lowest-order processes considered in our model, transitions between the degenerate ground-state configurations of triply occupied localized states are absent. Such transitions require higher-order processes beyond the present approximation. Consequently, no switching occurs within the scope of our model.

\newpage
\bibliography{reference}%

\end{document}